\documentclass[galaxies,article,accept,oneauthor,pdftex,10pt,a4paper]{mdpi} 

\firstpage{1} 
\articlenumber{x}
\doinum{10.3390/------}
\pubvolume{xx}
\pubyear{2016}
\copyrightyear{2016}
\history{Accepted by {\em Galaxies}}

 \usepackage{amssymb}
 \usepackage{textcomp}
 
 \theoremstyle{mdpi}
 \newcounter{thm}
 \newcounter{ex}
 \newcounter{re}
 \newcommand{\MHz}{\ensuremath{\textrm{ MHz}}}
 \newcommand{\GHz}{\ensuremath{\textrm{ GHz}}}
 \newcommand{\pc}{\ensuremath{\textrm{pc}}}
 \newcommand{\kpc}{\ensuremath{\textrm{kpc}}}
 \newcommand{\cm}{\ensuremath{\textrm{cm}}}
 \newcommand{\cucm}{\ensuremath{\textrm{cm}^{-3}}}
 \newcommand{\radmsq}{\ensuremath{\textrm{ rad m}^{-2}}}
 \newcommand{\uG}{\ensuremath{\mu\textrm{G}}}
 \newcommand{\K}{\ensuremath{\textrm{ K}}}
 
 \newcommand{\tphi}{\tilde{\phi}}
 
 \newcommand{\sinc}{\ensuremath{\mathrm{sinc}}}
 
 \newcommand{\Ha}{H$\alpha$}

\Title{Is there a polarization horizon?}

\Author{Alex S. Hill$^{1,2,3}$*}
\AuthorNames{Alex S. Hill}

\address{%
$^{1}$ \quad University of British Columbia, Vancouver, BC Canada; ashill@astro.ubc.ca \\
$^{2}$ \quad Space Science Institute, Boulder, CO USA \\
$^{3}$ \quad Dominion Radio Astrophysical Observatory, National Research Council, Penticton, BC Canada}

\abstract{Modern radio spectrometers make measurement of polarized intensity as a function of Faraday depth possible. I investigate the effect of depolarization along a model line of sight. I model sightlines with two components informed by observations: a warm ionized medium with a lognormal electron density distribution and a narrow, denser component simulating a spiral arm or H~{\sc ii} region, all with synchrotron-emitting gas mixed in. I then calculate the polarized intensity from $300-1800 \MHz$ and calculate the resulting Faraday depth spectrum. The idealized synthetic observations show far more Faraday complexity than is observed in Global Magneto-Ionic Medium Survey observations. In a model with a very nearby H~{\sc ii} region observed at low frequencies, most of the effects of a ``depolarization wall'' are evident: the H~{\sc ii} region depolarizes background emission and less (but not zero) information from beyond the H~{\sc ii} region reaches the observer. In other cases, the effects are not so clear, as significant amounts of information reach the observer even through significant depolarization, and it is not clear that low-frequency observations sample largely different volumes of the interstellar medium than high-frequency observations. The observed Faraday depth can be randomized such that it does not always have any correlation with the true Faraday depth.}


\begin{document}






\section{Introduction}


Polarized radio continuum emission in the Milky Way typically originates as synchrotron emission caused by relativistic particles accelerating due to magnetic fields. As the polarized emission propagates through the magnetized, ionized component of the interstellar medium (ISM), it undergoes Faraday rotation. A number of physical processes can cause polarization vectors to add destructively, leading to depolarization \citep{Burn:1966ug,SokoloffBykov:1998,GaenslerDickey:2001}. \citet{UYANIKERLandecker:2003} introduced the ``polarization horizon'', a distance beyond which all emission is depolarized (typically due to a combination of depth and beam depolarization). To first order, the line of sight beyond the polarization horizon does not contribute to the observed polarized emission. The polarization horizon is wavelength, angular resolution, and direction dependent, and true to the metaphor of a horizon is not a solid wall beyond which we cannot see.

With the angular resolution of interferometers, such as the Canadian Galactic Plane Survey (CGPS) observations used by \citet{UYANIKERLandecker:2003}, depolarization is often evident as a qualitative change in the morphology of the polarized emission. At this meeting, a number of authors presented observations in which there is not clear evidence of a polarization horizon as a general concept. \citet{van-EckHaverkorn:2017} showed $115-178 \MHz$ LOFAR observations over a $\approx (5^\circ)^2$ field, identifying multiple Faraday depth components and schematically tying them to two nearby primarily-neutral regions within $\approx 300 \pc$. Thomson showed $300-480 \MHz$ data from the all-sky Galactic Magnetoionic Medium Survey (GMIMS) Low Band South (GMIMS-LBS) obtained with the CSIRO Parkes Radiotelescope. In his observations, there is clear depolarization associated with specific H{\sc ii} regions, which I call a ``depolarization wall''. \citet{Wolleben:2007fw} used depolarization in a $1.4 \GHz$ polarization survey to construct a model for Radio Loop~I as the interaction of two shells. \citet{SunLandecker:2015} used depolarization in GMIMS-HBN data to estimate the distance to the North Polar Spur. We also see clear evidence for depolarization in the $1280-1750 \MHz$ GMIMS High Band North (GMIMS-HBN) data, where the W4 superbubble reduces the intensity of Fan Region emission by $\approx 30\%$ \citep{HillLandecker:2017}. In this case, though depolarization clearly affects the polarized intensity and is associated with ionized gas at a discrete distance in a spiral arm, the depolarization does not appear to remove all information about the gas behind the depolarizing structure.
Because depolarization is purely a polarization effect, there is no corresponding change in the morphology of total intensity images at the same wavelengths.

In this contribution, I evaluate the significance of the polarization horizon by integrating increasing distances through a model ISM, assessing the extent to which the polarization horizon metaphor is useful both in terms of the polarized intensity and the measured Faraday depth. 
To be concrete, I focus my analysis on the GMIMS-HBN and GMIMS-LBS frequency ranges as examples of a relatively high frequency with relatively narrow $\lambda^2$ coverage and a relatively low frequency with relatively wide $\lambda^2$ coverage, although the analysis is more broadly applicable.



\section{Propagation and emission of polarized radiation}

Polarized radiation is emitted in the diffuse interstellar medium and propagates to an observer as
\begin{equation} \label{eq:P}
\mathcal{P} = \int_D^0 \epsilon(s) (\vec{B} \times \hat{s})^2 e^{2 i \psi(s, \lambda)} ds.
\end{equation}
Here $s=0$ is the observer and $s=D$ is the back of the emission region. The emissivity $\epsilon(s)$ is a function of the relativistic electron density and energy spectrum and the polarization angle at the observer is
\begin{equation}
\psi(s, \lambda) = \psi_0(s) + \phi(s) \lambda^2.
\end{equation}
Polarized radiation is emitted at the position $s$ with polarization angle $\psi_0$ and then undergoes Faraday rotation as it propagates from $s$ to the observer proportional to the Faraday depth,
\begin{equation} \label{eq:phi}
\phi(s) = 0.81 \int_s^0 \frac{n_e(s')}{\cucm} \frac{\vec{B}(s') \cdot d\vec{s'}}{\uG \, \pc} \textrm{ rad m}^{-2},
\end{equation}
where $n_e$ is the thermal (non-relativistic) electron density and $\vec{B}$ is the magnetic field. We measure the real and imaginary parts of $\mathcal{P}$ as the Stokes parameters $Q$ and $U$, respectively. For a slab of uniform density with a uniform magnetic field and uniform synchrotron emissivity $\varepsilon$, equation~\ref{eq:P} simplifies to \citep{Burn:1966ug}
\begin{equation} \label{eq:depth}
\mathcal{P} = \varepsilon \cdot \left(\frac{\sin \phi \lambda^2}{\phi \lambda^2}\right) e^{2i \psi},
\end{equation}
where $\phi$ and $\psi$ are the Faraday depth and polarization angle integrated through the entire slab.

In the simplest case -- that of a background source which emits radiation at a uniform angle followed by the radiation propagating through a medium which is uniform in both electron density and magnetic field strength and direction -- this propagation results in a simple observed polarization vector in which
\begin{equation}
\psi(\lambda) = \psi_0 + \phi(\mathrm{source}) \lambda^2 \equiv \psi_0 + \mathrm{RM} \lambda^2.
\end{equation}
With observations at as few as two frequencies (though more measurements reduce ambiguities), the rotation measure RM, equivalent to the Faraday depth all the way to the emitting source, and intrinsic polarization angle are easy to derive as the slope and zero-intercept of the observed polarization angle as a function of $\lambda^2$.

\citet{Burn:1966ug} recognized that with measurements of $\mathcal{P}$ at many frequencies, one can disentangle the signature of multiple components along the line of sight. This is a computationally intensive process which did not become feasible until recently; \citet{BrentjensBruyn:2005} recognized the power of this technique and coined the term ``Faraday tomography.'' In reality, the ISM is inhomogeneous in $n_e$, $\vec{B}$, and cosmic ray electron density. The GMIMS survey \citep{WollebenLandecker:2009}, at $300 \MHz < \nu < 1750 \MHz$, as well as observations with LOFAR and MWA at lower frequencies, exploit the Faraday tomography technique to construct image cubes of diffuse polarized emission as a function of Faraday depth, providing three-dimensional information about the structure of the magnetized ISM. However, interpretation of these three-dimensional image cubes is complicated by depolarization effects.

In common situations, polarization vectors destructively interfere, causing depolarization. In {\em bandwidth depolarization}, there is sufficient change in $\lambda^2$ within a single channel so that $\Delta \psi = \phi \Delta \lambda^2 \gtrsim 1$ and the channel is depolarized. Here, I consider observations with sufficiently narrow channel spacing to avoid bandwidth depolarization. In {\em geometrical depolarization}, emission components along the line of sight have differing $\psi_0$. This mechanism is wavelength-independent. I define the models in this paper (\S~\ref{sec:methods} below) with constant $\psi_0$, so there is no geometrical depolarization. Note that geometrical depolarization is observed in {\em Planck} observations at wavelengths where Faraday rotation is negligible \cite{Miville-DeschenesYsard:2008,DelabrouilleBetoule:2013}. {\em Depth depolarization} is similar to geometrical depolarization but is due to Faraday rotation along the line of sight, causing more distant emission to reach the observer with a different polarization angle than closer emission, accounted for in the uniform case by the $\sinc$ term in parentheses in equation~\ref{eq:depth}. Changes in $\phi$ along the line of sight would similarly lead to depolarization, but equation~\ref{eq:depth} does not capture these effects. Depth depolarization is wavelength-dependent and thus can occur in the models presented here. In {\em beam depolarization}, emission with differing polarization vectors is present within the beam of the telescope, leading to destructive interference. Most commonly, the polarization angle varies within the beam due to individual sightlines within the beam with different amounts of Faraday rotation. This effect depends strongly upon the size of the beam and the observing wavelength. Beam depolarization by a slab with standard deviation of Faraday depths $\sigma_\phi$ is often parameterized by the depolarization factor \citep{Burn:1966ug},
\begin{equation} \label{eq:Burnp}
p \equiv \frac{L(\lambda)}{L(0)} = \exp \left(-2\sigma_\phi^2 \lambda^4 \right).
\end{equation}
This applies when $\sigma_\phi$ has a Gaussian distribution and for background polarized emission by a foreground depolarizing slab; the depolarization effects are presumably less severe (meaning $p$ is larger) when the emission is mixed with the slab. \citet{Tribble:1991us} points out that this approximation only holds when $p \gtrsim 0.5$; at longer wavelengths, the depolarization is less severe,
\begin{equation} \label{eq:Tribblep}
p = \frac{N^{-1/2}}{\sigma_\phi \lambda^2 2\sqrt{2}},
\end{equation}
where $N$ is the number of independent sightlines within the beam. In reality, the effects of different types of depolarization cannot be fully decoupled.

\begin{figure}[tbp]
\centering
\includegraphics[width=\textwidth]{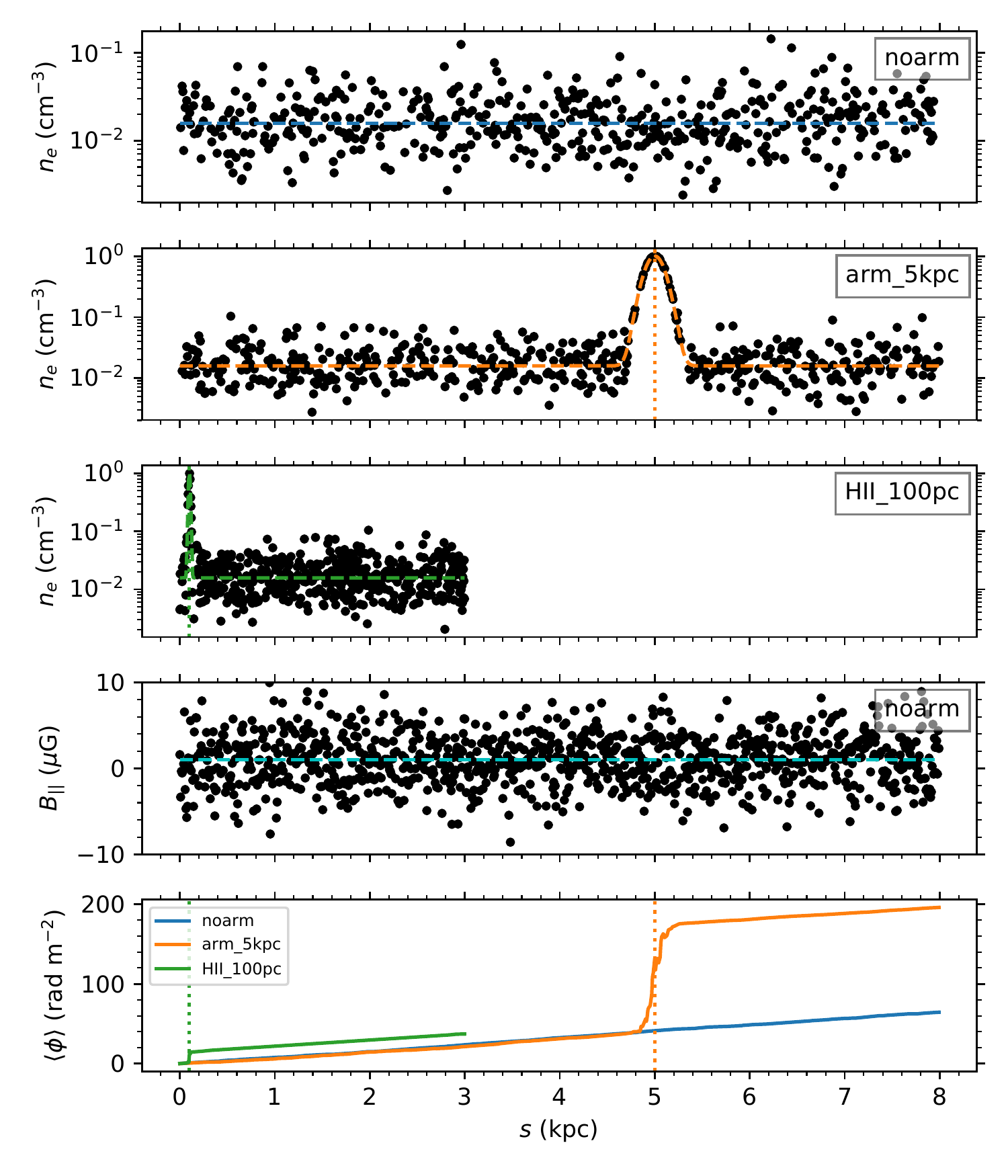}
\caption{Input density along the line of sight for one sightline in each model, with parameters listed in Table~\ref{tbl:inputs}. The ``beam'' in a given model consists of $N$ sightlines, each chosen from the same distributions as shown here. The fourth panel shows the line-of-sight component of the magnetic field for one model. Points show the individual cells; dashed lines show the median. Vertical dotted lines show the position of the arm in each model. The bottom panel shows the mean Faraday depth across all the sightlines in each model.}
\label{fig:inputs}
\end{figure}

\section{Methods} \label{sec:methods}

Based on observations of pulsar dispersion and H$\alpha$ emission, we have reasonable constraints on the electron density distribution in the ISM. The warm ionized medium (WIM), which contains most of the mass of ionized hydrogen in the Galaxy \citep{HaffnerDettmar:2009} and is therefore presumably where most Faraday rotation occurs, is turbulent \citep{Benjamin:1999jx} and transsonic \citep{HillBenjamin:2008,BerkhuijsenFletcher:2008,GaenslerHaverkorn:2011}. Therefore, a series of compressions and rarefactions establishes the electron density structure. This naturally leads to a lognormal distribution of electron density \citep{Vazquez-Semadeni:1994,PassotVazquez-Semadeni:1998} which is also inferred from the integrated observables, H$\alpha$ emission measure $\mathrm{EM} = \int n_e^2 \, ds$ and pulsar dispersion measure $\mathrm{DM} = \int n_e \, ds$ \citep{HillBenjamin:2008,BerkhuijsenFletcher:2008}.

I have constructed simple numerical models of depolarization of propagating radio waves with inputs listed in Table~\ref{tbl:inputs}. I begin by assuming a distribution of thermal electron density based on observations of \Ha\ emission and pulsar dispersion. Specifically, I assume a lognormal distribution of $n_e$ and that the ionized gas fills a fraction $f$ of the line of sight. I set $f=0.5$, consistent with modelling of H$\alpha$ and pulsar dispersion observations at latitudes $|b| > 10^\circ$ \citep{HillBenjamin:2008}, so half of the cells (chosen randomly) have $n_e = 0$ and the densities in the other half are drawn from the lognormal distribution. To this distribution, I add a term modelling a denser region which I refer to as a spiral arm (though it can also be thought of as an H~{\sc ii} region depending on the chosen values for $n_\mathrm{arm}$ and $w_\mathrm{arm}$), so 
\begin{equation}
n(s) = n_i +  n_\mathrm{arm} \exp\left( -\frac{(s - d_\mathrm{arm})^2}{2w_\mathrm{arm}^2} \right)
\end{equation}
where $n_i$ is drawn from a lognormal distribution with most probable value $10^\mu$ and standard deviation $\sigma_{\log n}$. I construct one model with no arm, one model with an arm $5 \, \kpc$ from the observer, and one model with an ``arm'' that is very nearby ($100 \, \pc$) and is more like a nearby, low-density H~{\sc ii} region. I also assume a magnetic field geometry loosely informed by models of the Galactic magnetic field \citep{SunReich:2008,JanssonFarrar:2012}. The total magnetic field $|\vec{B}|$ is drawn from a Gaussian distribution with mean $B_0$ and standard deviation $\sigma_B$; the line of sight component is $B_{||} = |\vec{B}| \sin \theta$. I further assume a uniform cosmic ray electron density and emissivity $\varepsilon(\lambda=20 \, \cm)= 1 \K / \kpc$ with temperature spectral index $\beta = 3.2$. I chose a constant $\psi_0 = 45^\circ$ for simplicity, although this is inconsistent with the non-uniform magnetic field. Figure~\ref{fig:inputs} shows $n_e$ and $B_{||}$ along one sightline as well as the resulting mean Faraday depth $\phi$ as a function of distance in the three models.

\begin{figure}[tb]
\centering
\includegraphics[width=\textwidth]{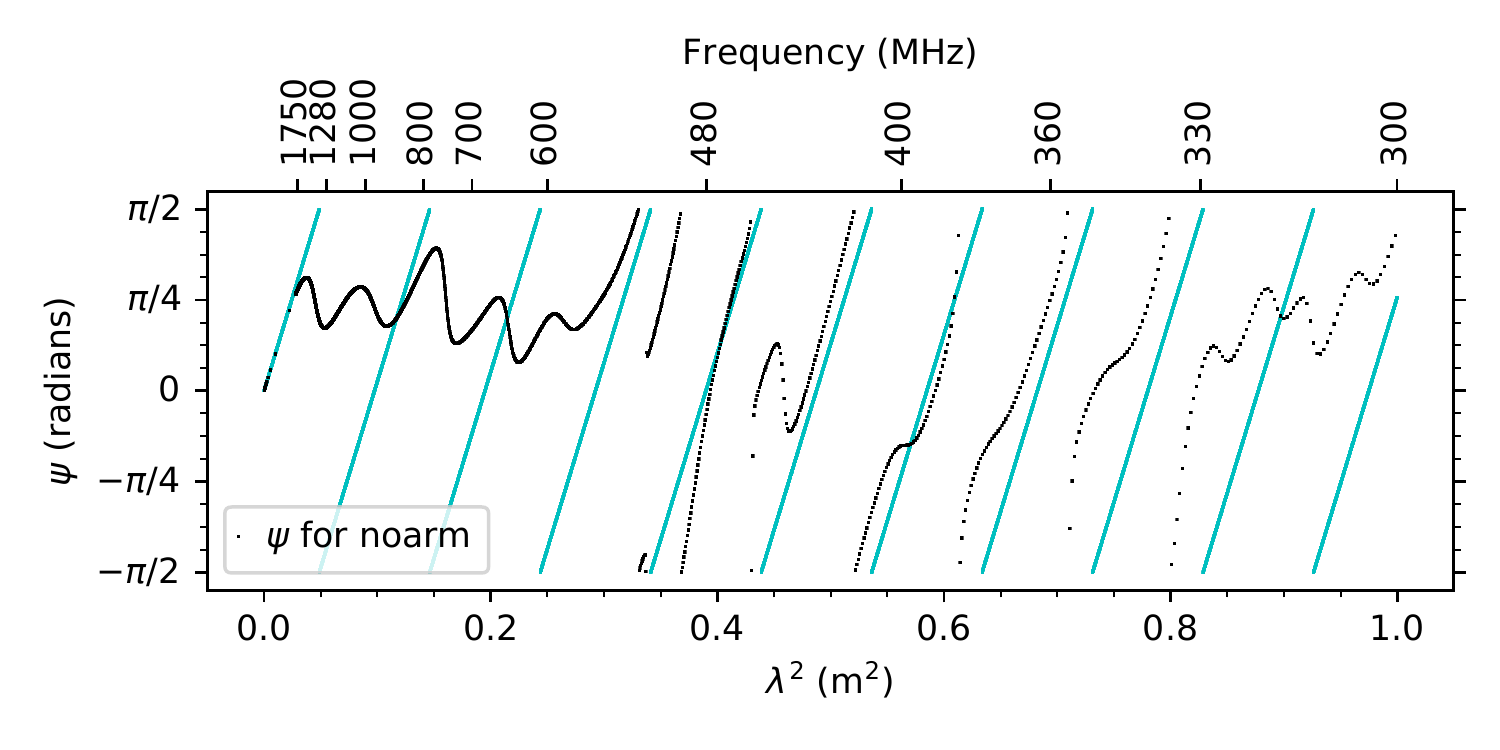}
\caption{Spectrum of polarization angle as a function of wavelength integrated over a beam (the sum of $N=25$ sightlines) all the way through the {\tt noarm} model, calculated on $0.5 \MHz$ intervals. Cyan lines show the slope $d\psi/d\lambda^2 = \phi / 2$, as expected for a uniform slab, where $\phi = 37.1 \textrm{ rad m}^2$ is the mean Faraday depth integrated through all sightlines.}
\label{fig:spectrum}
\end{figure}

With these inputs, I integrate equations~\ref{eq:P}--\ref{eq:phi} numerically to calculate the observed polarization vector $\mathcal{P}(\lambda)$ on $0.5 \MHz$ intervals. I run this calculation for independent sightlines and, to simulate the effects of beam depolarization, combine $\mathcal{P}(\lambda)$ from $N$ sightlines. I experimented with values of $N$ up to 100 but found that $N=25$ is sufficiently large for the qualitative behavior to be similar to higher-$N$ models and sufficiently small for the three models to fit in memory on a laptop computer. The linearly polarized intensity for the beam is then $L = \left| \left\langle \mathcal{P}_i \right\rangle \right| = (\langle Q_i \rangle^2 + \langle U_i \rangle^2)^{1/2}$, where the angle brackets denote averaging over the $N$ sightlines, while the polarized intensity for an individual sightline is $L_i = | \mathcal{P}_i |$. I show the spectrum of $\psi = 0.5 \arctan(U/Q)$ for one model in Figure~\ref{fig:spectrum}. From this spectrum, I perform RM synthesis \citep{BrentjensBruyn:2005} and RM cleaning \citep{Heald:2009kc} using the {\tt pyrmsynth} package\footnote{\url{https://github.com/mrbell/pyrmsynth}} to calculate the Faraday depth spectrum. I performed RM synthesis in the two GMIMS frequency ranges, but not accounting for radio frequency interferences (RFI), other excluded frequency channels, noise, or any other instrumental realities. To distinguish the Faraday depth measured from RM synthesis from the true Faraday depth (which is known precisely as a function of $s$ in these models but not in the real ISM), I use the notation $\tphi$ to denote the Faraday depth derived from RM synthesis. The input frequency ranges and output maximum value of $\tphi$, maximum scale of Faraday depth features, FWHM of the RM transfer function, and the range and chosen resolution $\delta \tphi$ of calculated Faraday depth spectra are listed in Table~\ref{tbl:RMsynthesis}, with these values calculated following \citet{SchnitzelerKatgert:2009}.

\newcommand{\tblhd}[1]{\multicolumn{1}{c}{\textbf{#1}}}

\begin{table}[tb]
\caption{Input parameters}
\label{tbl:inputs}
\small 
\centering
\begin{tabular}{l rrrrrrrrrrr}
\toprule
\tblhd{Name}	& \tblhd{$10^\mu$}	& \tblhd{$\sigma_{\log n}$} & \tblhd{$f$} & \tblhd{$n_\mathrm{arm}$} & \tblhd{$d_\mathrm{arm}$} & \tblhd{$w_\mathrm{arm}$} & \tblhd{$\mathrm{DM}$} & \tblhd{$B_0$} & \tblhd{$\sigma_B$} & $\theta$ & $N$ \\
 & (cm$^{-3}$) & & & (cm$^{-3}$) & (pc) & (pc) & (pc cm$^{-3}$) & ($\mu$G) & ($\mu$G) & &  \\
\midrule
{\tt noarm     } & $0.016$ & $0.30$ & $0.5$ & $0.0$ & ...    & ...   & $78.6$ & $1.0$ & $3.0$ & $45^\circ$ & $25$ \\
{\tt arm\_5kpc } & $0.016$ & $0.30$ & $0.5$ & $1.0$ & $5000$ & $100$ & $204.8$ & $1.0$ & $3.0$ & $45^\circ$ & $25$ \\
{\tt HII\_100pc} & $0.016$ & $0.30$ & $0.5$ & $1.0$ & $100$  & $10$  & $42.2$ & $1.0$ & $3.0$ & $45^\circ$ & $25$ \\
\bottomrule
\end{tabular}
\end{table}

\begin{table}[tb]
\caption{RM synthesis parameters}
\label{tbl:RMsynthesis}
\small
\centering
\begin{tabular}{l r@{$-$}l r c rrrrr }
\toprule
\tblhd{Survey} & \multicolumn{2}{c}{Frequencies} & \tblhd{$\delta \nu$} & & \tblhd{$\tphi_\mathrm{max}$} & \tblhd{Max scale} & \tblhd{FWHM} & \tblhd{Range} & \tblhd{$\delta \tphi$} \\
\cline{2-4}
\cline{6-10}
\tblhd{} & \multicolumn{3}{c}{(MHz)} && \multicolumn{5}{c}{(rad m$^{-2}$)} \\
\midrule
GMIMS-LBS & $300$&$480$ & $0.5$ && $570$ & $8.0$ & $6.2$ & $-300 < \tphi < +300$ & $1$ \\
GMIMS-HBN & $1280$&$1750$ & $0.5$ && $44 \, 000$ & $107$ & $149$ & $-600 < \tphi < +600$ & $5$ \\
\bottomrule
\end{tabular}
\end{table}

By constructing independent sightlines, I neglect projection effects and any correlation between nearby zones at a given distance. Moreover, the assumption that $B$ and $n_e$ are uncorrelated is unlikely to be true in reality. The Faraday depth in this model is the integral of the product of a Gaussian ($B$) and a lognormal ($n_e$) distribution and therefore is itself neither Gaussian nor lognormal. Therefore, the assumption of a Gaussian distribution in the derivation of equations~\ref{eq:Burnp} and \ref{eq:Tribblep} does not apply. The fixed $\psi_0$ is inconsistent with independent sightlines. I also treat all sightlines as in the plane of the Galaxy, neglecting the different vertical distributions of the thermal electrons, relativistic electrons, and magnetic fields as well as reversals or other changes in the mean magnetic field along the sightline; \citet{SokoloffBykov:1998} address these cases. Because this model assumes sightlines in the midplane but the $f=0.5$ assumption is based on modelling of sightlines which sample the entire WIM disk, the filling fraction may be overestimated. The effect of this would simply be for the numerical values of the distances in this paper to be lower than the distances they model in the real Galaxy.

\section{Results} \label{sec:results}


The spectrum in Figure~\ref{fig:spectrum} shows the impact of Faraday rotation and depolarization effects. First, the different $\lambda^2$ coverages of the GMIMS-LBS ($300-480 \MHz$) and GMIMS-HBN ($1280-1750 \MHz$) surveys are evident from the top axis. In the GMIMS-LBS band, the polarization angle rotates many times through $\pi$ radians. 
At frequencies $\nu \gtrsim 500 \MHz$, there are no more $\pi$ radian rotations in the polarization angle. Instead, $\psi$ oscillates on a period of $\approx 0.05 \textrm{ m}^2$. There is always a limit in which a spectral window is narrow enough so that the data are consistent with a straight line (even if it is sampled at a sufficient number of frequencies to avoid the $n\pi$ ambiguity); in such a narrow window, a single rotation measure would describe the data fully. We therefore must consider whether our $\lambda^2$ coverage is sufficient to be comfortably out of that limit; this is one definition of ``wide'' $\lambda^2$ coverage. Most windows with the $\lambda^2$ coverage of the GMIMS-HBN data are narrow enough to be in this situation. Moreover, both the magnitude  and the sign of the derived rotation measure in these windows --- especially at $\nu \gtrsim 600 \MHz$ in the example in Figure~\ref{fig:spectrum} --- can change depending on the chosen spectral window: these windows may contain insufficient variation in $\psi$ for the RM synthesis process to identify even the correct sign of $\phi$. Therefore, {\em it is plausible and even likely that the sign of the Faraday depth derived from emission over a narrow $\lambda^2$ range in a given observation will disagree with the sign of the rotation measure towards a background point source}. In this situation, which is also evident in Figures~2 and 4 of \citeauthor{SokoloffBykov:1998} \citep{SokoloffBykov:1998,SokoloffBykov:1999}, $\tphi$ could have a negative sign even though there is no volume along the sightline in which $\phi$ is negative. The highest frequency at which $\pi$ radian rotations occur is $\approx 650 \MHz$ in the {\tt arm\_5kpc} model, and in the {\tt HII\_100pc} model, there are $\pi$ radian rotations down to $\lambda^2 = 0$, so this behavior is dependent upon the depolarizing structures in the ISM.

\begin{figure}[tb]
\centering
\includegraphics[width=\textwidth]{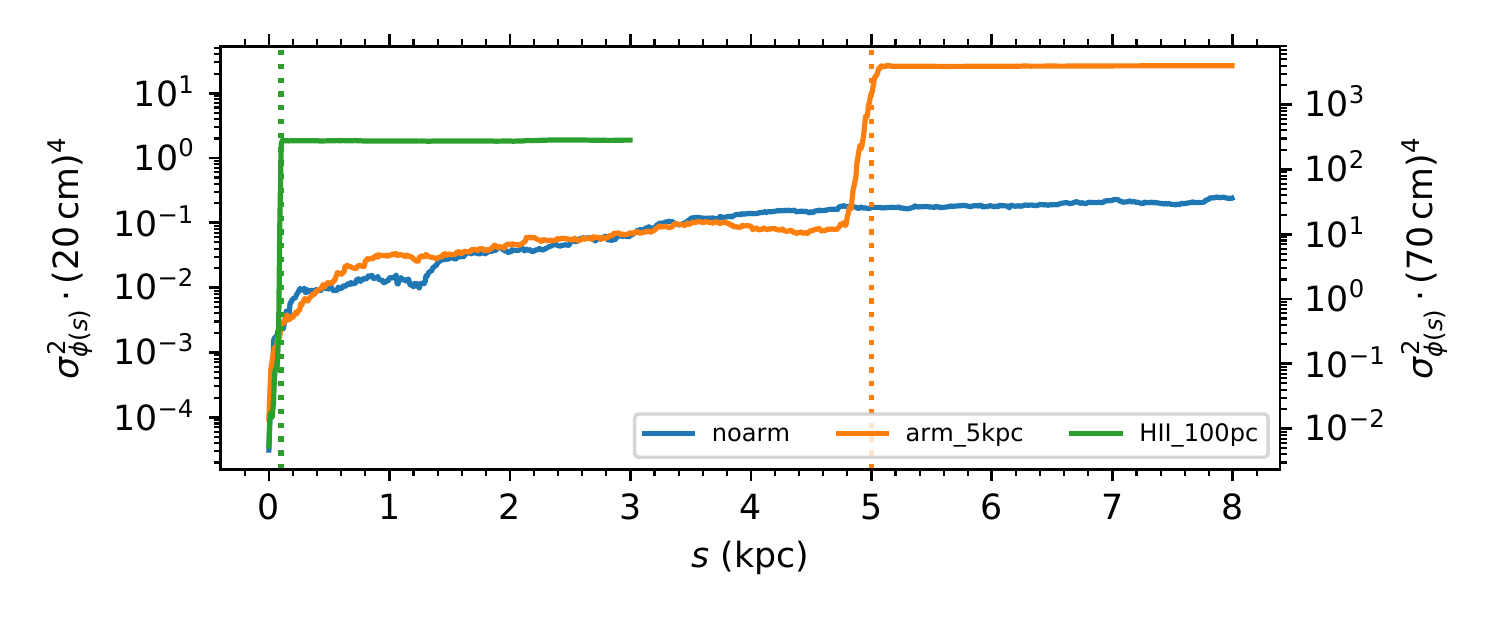}
\caption{The factor in equation~\ref{eq:Burnp}, $\sigma_\phi^2 \, \lambda^4$, for each of the models. The left axis values are labeled for $\lambda = 20 \, \cm$; the right axis values are labeled for $\lambda = 70 \, \cm$. When $\sigma_\phi^2 \lambda^4 \gg 10^0$, total depolarization is expected. Dotted vertical lines show the position of the arm in each model.}
\label{fig:siglambda}
\end{figure}

If equation~\ref{eq:Burnp} applies, we expect beam depolarization to cause the polarization horizon to be approached when $\sigma_\phi^2 \lambda^4 \gtrsim 1$. This is shown in Figure~\ref{fig:siglambda} at two wavelengths, $\lambda = 70 \, \cm$ (in the GMIMS-LBS band) and $\lambda=20 \, \cm$ (in the GMIMS-HBN band). At $20 \, \cm$, this threshold is never reached in the {\tt noarm} model but is reached in the spiral arm at $s = 5 \kpc$ in model {\tt arm\_5kpc} and (marginally) in the H~{\sc II} region at $s=100 \, \pc$ in model {\tt HII\_100pc}. At $\lambda = 70 \, \cm$, this threshold is reached in the H~{\sc ii} region in that model and by $s \approx 1 \, \kpc$ in the other two models. Therefore, one expects the polarization horizon in the long-wavelength case to be quite nearby. At the other extreme, beam depolarization is not likely to cause a polarization horizon at the shorter wavelength at all in the {\tt noarm} model, although other depolarization effects such as variations in $\psi_0$ could.

\begin{figure}[tbp]
\centering
\includegraphics[width=0.85\textwidth]{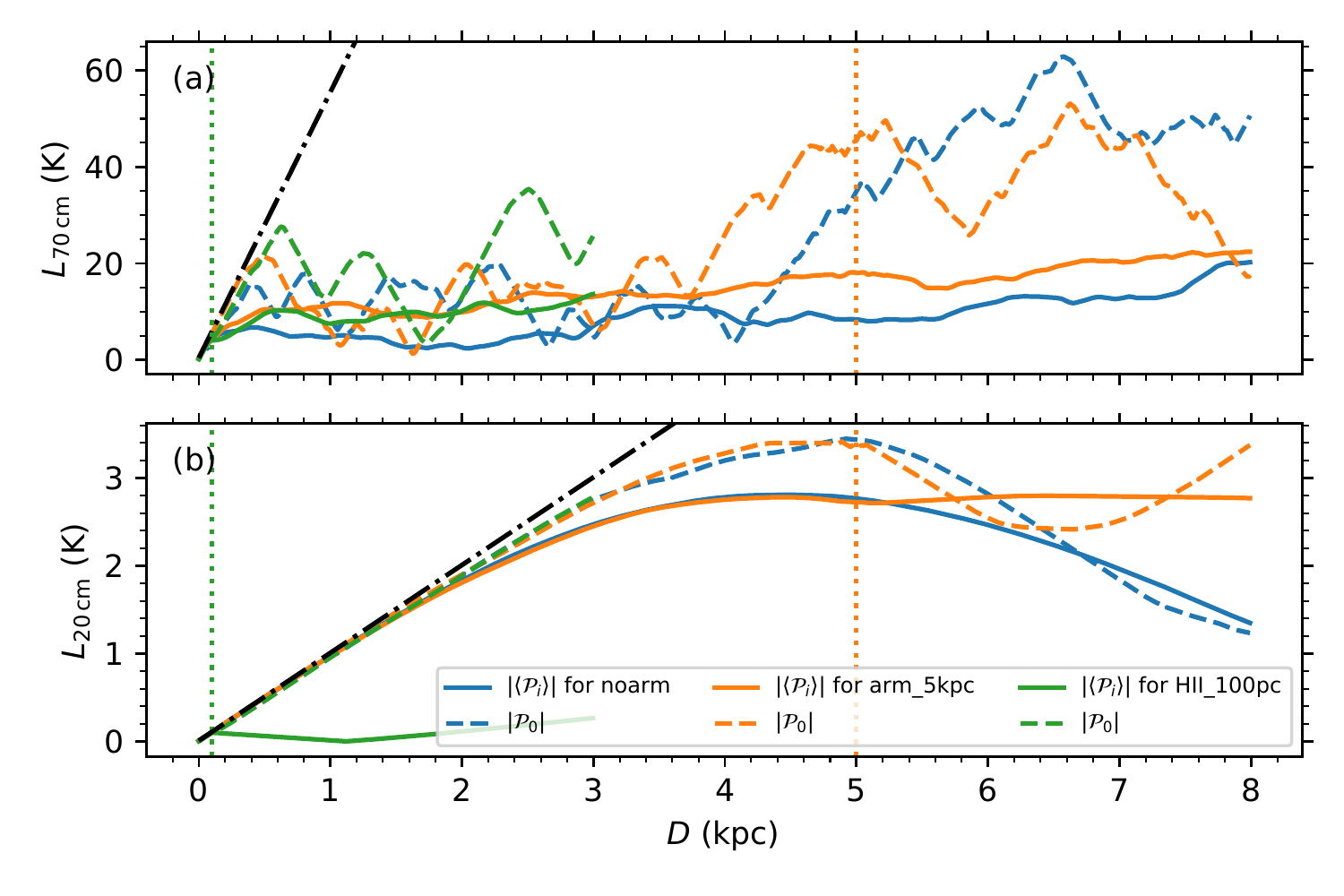}
\caption{Polarized intensity as a function of $D$, the integration limit in equation~\ref{eq:P}, at $\lambda = 70 \, \cm$ (top) and $20 \, \cm$ (bottom). The solid lines show the polarized intensity over the beam, $| \langle \mathcal{P}_i \rangle |$. The dashed lines show the polarized intensity in one randomly chosen individual sightlines within the beam, $| \mathcal{P}_i |$. Vertical dotted lines show the position of the spiral arm in the model with the corresponding color. Dot-dashed lines show the polarized intensity in the absence of depolarization effects (equation~\ref{eq:PI}). Note that integrated polarized intensites (moment 0) are shown in Figure~\ref{fig:moments} below. Dotted vertical lines show the position of the arm in each model.}
\label{fig:results}
\end{figure} 

\begin{figure}[tbp]
\centering
\includegraphics[width=0.85\textwidth]{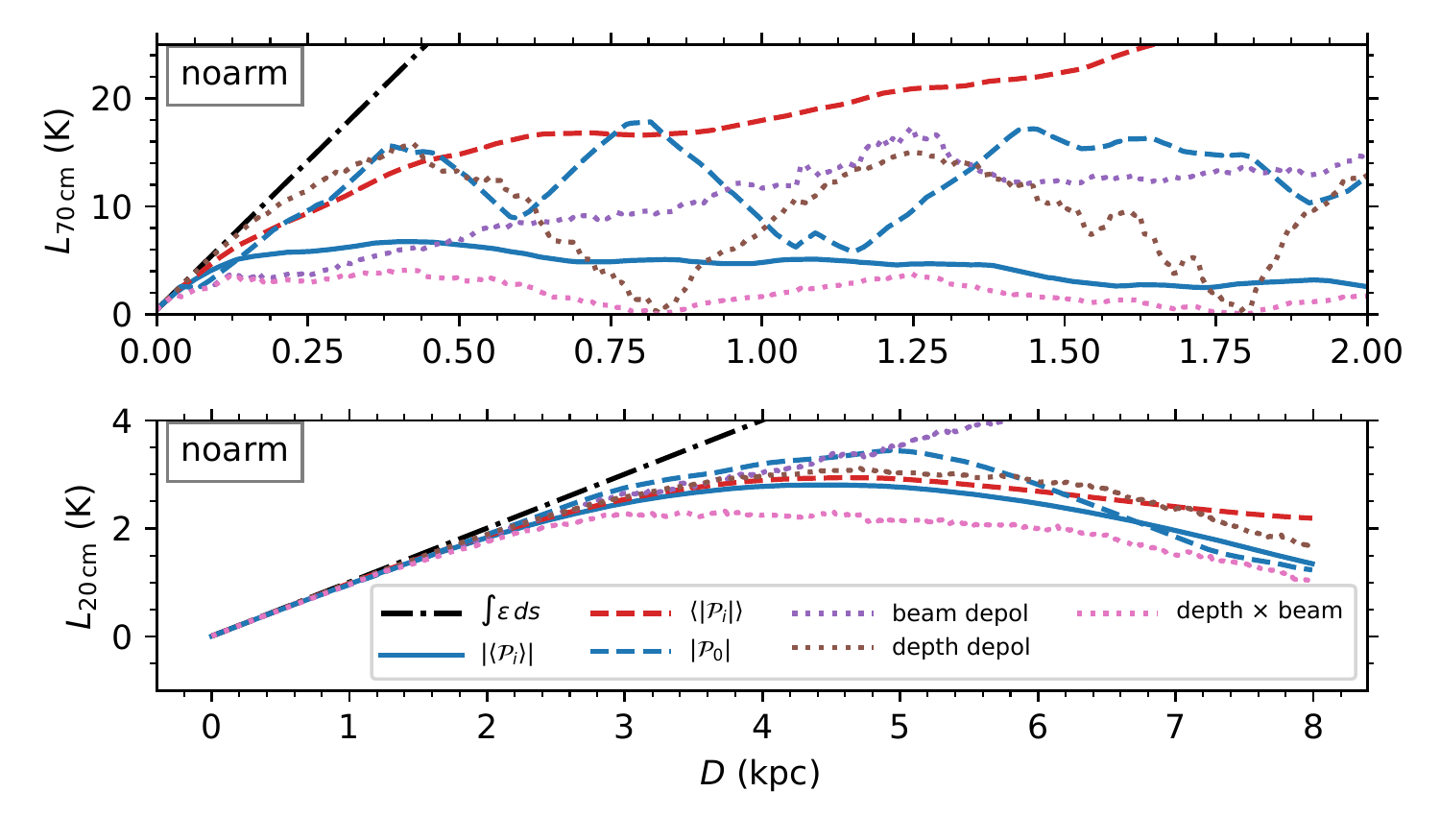}
\caption{Blue lines are as in Figure~\ref{fig:results}: polarized intensity for the full beam $|\langle \mathcal{P}_i \rangle|$ (blue solid line) and in a representative sightline within the beam $| \mathcal{P}_0 |$ (blue dashed line) from the {\tt noarm} model as a function of $D$. I also show the mean of $|\mathcal{P}_i|$, $\langle | \mathcal{P}_i | \rangle$, of the $N=25$ individual sightlines (red dashed line). Dotted lines show analytic models of beam depolarization (equations~\ref{eq:Burnp} and \ref{eq:Tribblep}), depth depolarization (equation~\ref{eq:depth}), and both. Note that in the top ($70 \, \cm$) panel, I show only the first $2 \, \kpc$ because depolarization effects become dominant on shorter distance scales at the long wavelength.}
\label{fig:depol_mechanisms}
\end{figure}

I show the output polarized intensity and polarization angle as a function of $D$ at the same two single wavelengths in each of the three models in Figure~\ref{fig:results}. The dashed lines show the polarized intensity of an individual sightline, which cannot include any beam depolarization effects. At $\lambda = 70 \, \cm$ (Fig.~\ref{fig:results}a), the polarized intensity increases and decreases with distance due to the $\sinc$ behavior of depth depolarization (Equation~\ref{eq:depth}). The solid lines show the entire beam, which does include beam depolarization effects in addition to depth depolarization. In all cases, in the absence of any depolarization, the polarized intensity would be
\begin{equation} \label{eq:PI}
L = \int_D^0 \varepsilon(s) ds = \varepsilon \, D \left( \frac{\lambda}{20 \textrm{ cm}} \right)^\beta
\end{equation}
for the inputs chosen in the model; this is shown with the black dot-dashed line. The polarized intensity in individual sightlines oscillates with distance and, after a distance such that $\phi(D) \lambda^2 \gtrsim 1$, is much lower than that in equation~\ref{eq:PI} due to depth depolarization. However, because $n_e$ and $B_{||}$ are not uniform, depth depolarization never produces the complete nulls implied by equation~\ref{eq:depth}.

In Figure~\ref{fig:depol_mechanisms}, I show the polarized intensity from the {\tt noarm} model compared to the predicted effects of depth (eq.~\ref{eq:depth}) and beam (eq.~\ref{eq:Burnp} and \ref{eq:Tribblep}) depolarization. At $\lambda = 20 \, \cm$ (bottom panel), the polarized intensity measured for the beam decreases from $100\%$ of the mean polarized intensity of the individual sightlines at $s=0$ to $\approx 75\%$ of that of the individual sightlines at $s = 8 \, \kpc$. The depth depolarization model (brown dotted line) accurately describes the mean of the individual sightlines (red dashed line); beam depolarization without considering depth depolarization (purple dotted line) does not. At this wavelength, the first depth depolarization null is not reached in $8 \kpc$. The combination of depth and beam depolarization (pink dotted line) produces a polarized intensity somewhat lower than seen in the simulation. This is unsurprising because equation~\ref{eq:Burnp} assumes that the emission is behind the depolarizing slab and that $\phi$ has a Gaussian distribution; neither assumption applies to the simulations.

At $\lambda = 70 \, \cm$, the first depth depolarization null is reached after $\approx 700 \, \pc$ (brown dotted line in Figure~\ref{fig:depol_mechanisms}, top panel), calculated using $\phi \lambda^2 = \pi$ from equation~\ref{eq:depth} with the mean value of $\phi$ across all sightlines. In a representative individual sightline (green dashed line), the polarized intensity comes close to zero but never fully reaches zero because $n_e$ and $B_{||}$ are not uniform. The mean polarized intensity across all sightlines (orange dashed line) never approaches zero because the locations of the nulls are out of phase in the different sightlines. The polarized intensity measured for the beam (blue solid line) is much smaller than the mean polarized intensity for the individual sightlines, indicating strong beam depolarization. However, equations~\ref{eq:Burnp} and \ref{eq:Tribblep} again overestimate the effects of beam depolarization; the polarized intensity across the beam is a factor of $>2$ larger than predicted (purple dotted line) for most values of $D$.

\begin{figure}[tb]
\centering
\includegraphics[width=\textwidth]{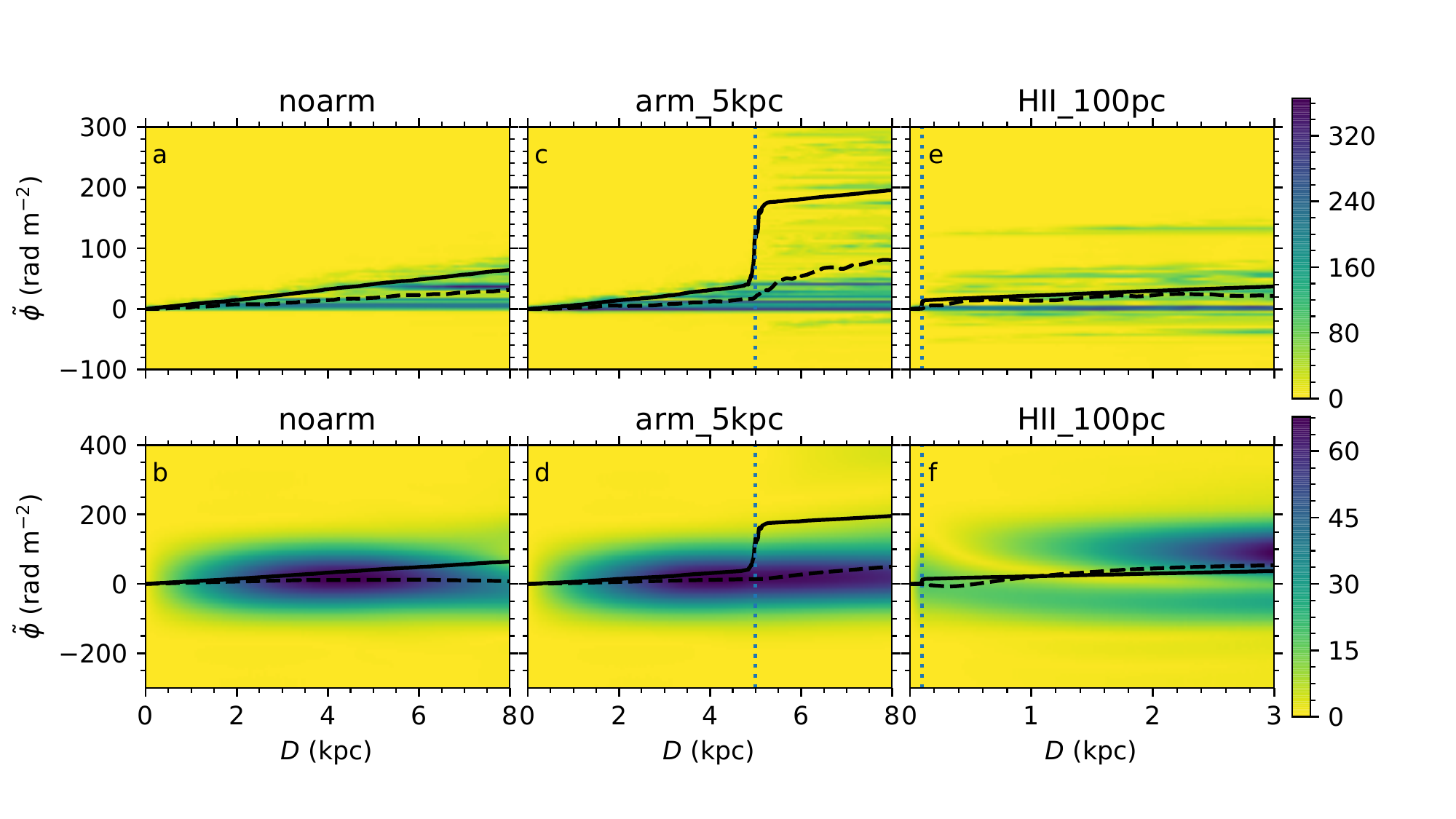}
\caption{Faraday depth spectrum as a function of $D$ as measured in the GMIMS-LBS frequency band (top) and the GMIMS-HBN frequency band (bottom). The simulated polarized intensity in K is indicated with the color bars. Solid lines show $\phi$ integrated to $s=D$; dashed lines show the first moment of the Faraday depth spectra. Dotted vertical lines show the position of the arm in each model.}
\label{fig:fdf}
\end{figure}

\begin{figure}[tb]
\centering
\includegraphics[width=\textwidth]{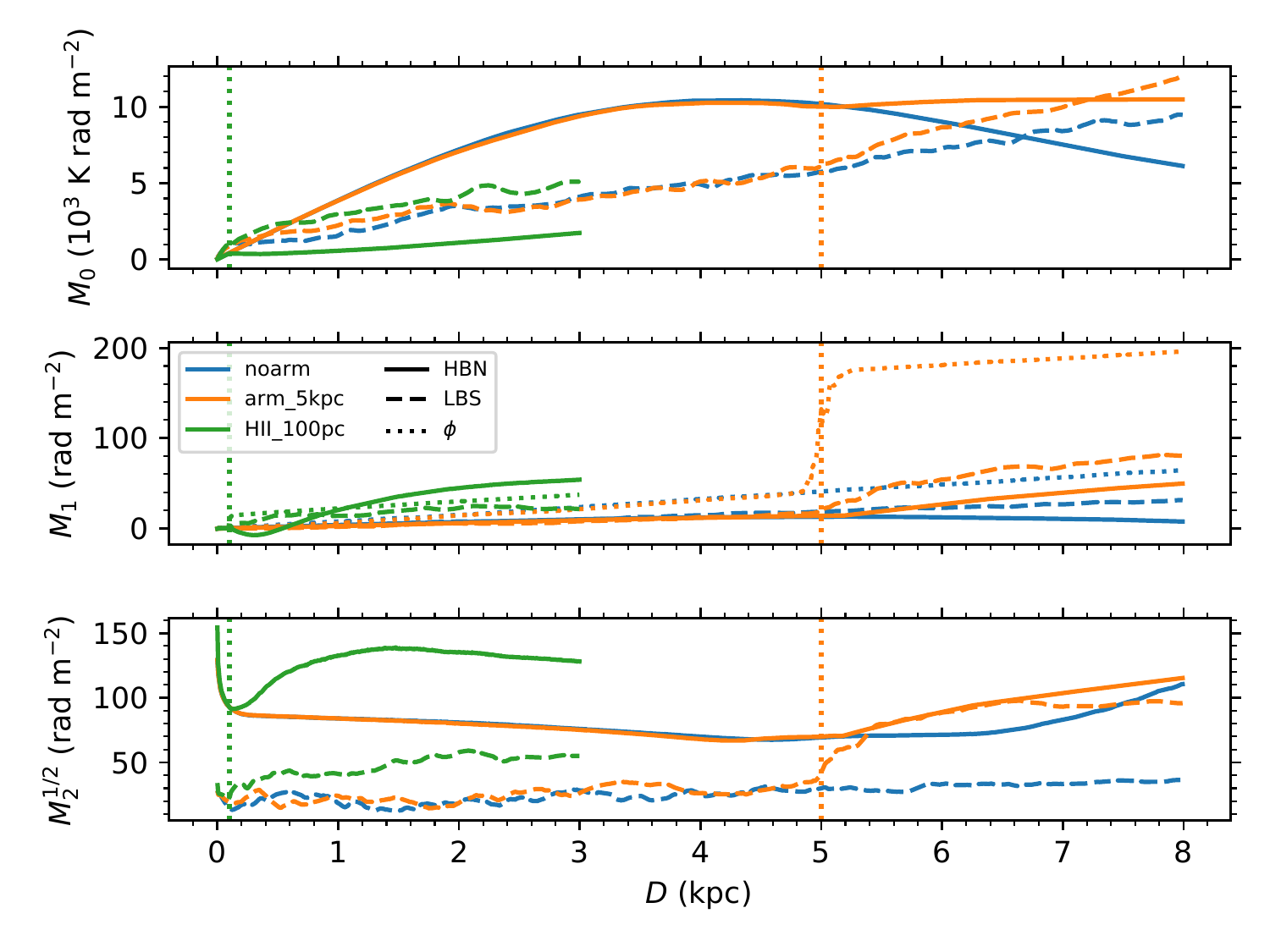}
\caption{Moments of the Faraday depth spectrum as a function of distance along the sightline. Solid lines show the GMIMS-HBN Faraday spectrum; dashed lines show the GMIMS-LBS Faraday spectrum. The zeroth moment is the polarized intensity integrated across the Faraday spectrum, $M_0 = \int L \, d\phi$. The first moment is the intensity-weighted Faraday depth, $M_1=\int L \phi \, d\phi/M_0$; the dotted lines show the true $\phi$ for the input model (from Figure~\ref{fig:inputs}), which is in principle the rotation measure one would measure towards a point source at distance $D$ (or a background point source for the maximum value of $D$). The second moment is the intensity-weighted width of the Faraday depth spectrum, $\int L \cdot (\phi - M_1)^2 d\phi/M_0$; $M_2^{1/2}$ is shown. Dotted vertical lines show the position of the arm in each model.}
\label{fig:moments}
\end{figure}

In Figure~\ref{fig:fdf}, I show the effect of depolarization on the polarized intensity measured using RM synthesis. In the {\tt noarm} model, as one integrates further along the line of sight to emission regions at higher Faraday depths, the Faraday spectrum has power at Faraday depths ranging from zero to $\phi(D)$, most clearly in the LBS band (Figure~\ref{fig:fdf}a). In the {\tt arm\_5kpc} model, the effects of the arm are evident in Figure~\ref{fig:fdf}c,d in both frequency ranges: the arm shifts some of the emission to much higher $\tphi$, including Faraday depths much higher than the mean $\phi(5 \, \kpc)$. However, there is little impact on the {\em total} polarized intensity either at a single frequency (Figure~\ref{fig:results}) or in the Faraday depth spectra (Figure~\ref{fig:fdf}) or moment 0 (Figure~\ref{fig:moments}).

Figure~\ref{fig:moments} shows moments of Faraday depth \citep{DickeyLandecker:2018}, analogous to moments of spectral lines. Moment 0 shows the polarized intensity integrated across all Faraday depths, moment 1 shows the polarized intensity-weighted mean Faraday depth, and moment 2 shows the intensity-weighted width of the Faraday depth spectra. From the moment 0 panel, it is clear that the LBS polarized intensity (dashed lines) increases approximately monotonically with distance in all models, even beyond the depolarizing features.

\section{Caveats}

In this work, I describe the Faraday spectra from individual realizations of a beam. Given the parameters I chose, the details vary significantly in different realizations. A detailed description of the possible spectra for a given set of parameters is beyond the scope of this work, as is an exploration of the many free parameters such as the chosen position of the spiral arms, the density distribution, and variations with Galactocentric radius, Galactic longitude, and latitude. Instead, I attempt to give a qualitative understanding of the implications of this model. I have tested many different realizations of the model, and the general findings hold. I have also tried different parameter choices; the three models I chose to present are representative examples of three different cases which highlight distinct effects in the GMIMS-HBN and GMIMS-LBS frequency ranges.

It is also not clear that the synthetic Faraday spectra presented here look anything like any observed Faraday spectra; they do not look like the Faraday spectra seen in GMIMS observations, which even in GMIMS-LBS data do not typically show the many components evident in Figure~\ref{fig:fdf}a,c,e (M.\ Wolleben et al. in prep). However, LOFAR Faraday depth spectra presented by \citet[and at this meeting]{van-EckHaverkorn:2017} do show notably complex spectra. Due primarily to RFI, the spectral sampling in GMIMS data is intermittent, which will likely map to incomplete sampling of Faraday depth space. Moreover, the models presented here do not include any noise; in a real telescope, noise may limit sensitivity to Faraday complexity. These models provide insight into the behavior of polarized radiation in the ISM, but it is not immediately obvious how to map this information onto real observations. Testing these models with observations will require generating many realizations of the models to generate statistics and synthetic images as seen only at the observer. In particular, one could use an extended region derived from many realizations of the {\tt noarm} model as a background and add a region with a radius of several beams from the {\tt HII\_100pc} model to see if the resulting image is qualitatively similar to observed depolarizing screens \citep[and Thomson's work at this meeting]{UYANIKERLandecker:2003,LandeckerReich:2010}.

\section{Discussion: Is there a polarization horizon?}

In the models presented here, depolarization acts on the single-frequency polarized intensity largely in accordance with the models by \citet{Burn:1966ug}, \citet{Tribble:1991us}, and \citet{SokoloffBykov:1998}. The behavior is qualitatively different at $\lambda = 70 \, \cm$ than at $20 \, \cm$: at the longer wavelength, beam depolarization effects due to the diffuse ISM dominate within a few hundred pc, before depth depolarization effects become important, while at $20 \, \cm$, depth depolarization is more important than beam depolarization with 25 sightlines within the beam.

There is a distance beyond which depolarization effects become evident. Especially in model {\tt HII\_100pc} observed at long wavelengths, that distance is very short, as predicted by the \citet{Burn:1966ug} model: $\sigma_\phi^2 \lambda^4 \gtrsim 1$ at $s \gtrsim 90 \, \pc$. In that most extreme case, integrating further beyond the depolarizing structure leads to relatively little change in the polarized intensity at $70 \, \cm$.  However, although the polarized intensity at the single frequency shown no longer increases with distance, the Faraday spectrum (Figure~\ref{fig:fdf}e) and the moments (green dashed lines in Figure~\ref{fig:moments}) do continue to evolve. This is the closest case in these models to a ``depolarization wall'', but the wall is not opaque to polarized radiation. Similarly, in the other models and in both frequency ranges, there is no physical distance beyond which further integration does not affect at least some aspects of the observed spectrum. Therefore, though it is probably reasonable to say that there is a different (but difficult to precisely define) weighting function to the volume sampled by polarization observations at different frequencies (with longer-wavelength observations weighted more towards a more nearby volume, especially in the {\tt HII\_100pc} model), these models suggest that longer wavelengths do not sample an entirely different volume than the shorter wavelengths.

In the same model observed at longer wavelengths, again the polarized intensity changes, first decreasing until $D \approx 1.1 \, \kpc$ and then increasing, presumably due to depth depolarization effects (green line in Figure~\ref{fig:results}b). However, the Faraday depth spectrum changes considerably as we integrate further (Figure~\ref{fig:fdf}f). In fact, the Faraday depth spectrum bifurcates into two components, with one at $\tphi \approx -60 \radmsq$ and one at $\tphi \approx +90 \radmsq$ even though there is no emission component at {\em either} Faraday depth; all of the gas in the model is found at $0 \lesssim \phi \lesssim +40 \radmsq$.

The situation is rather less hopeless for the first Faraday moments shown in Figure~\ref{fig:moments} and as dashed lines in Figure~\ref{fig:fdf} than for the individual components. Indeed, the first moments of $\tphi$ are generally close to $\phi/2$ for the models in which the emission and rotation are mixed, as expected for a uniform slab. Moreover, in the {\tt HII\_100pc} model (Figure~\ref{fig:fdf}e,f), the first moment of $\tphi$ is approximately $\phi$. This is expected because most of the Faraday rotation occurs in the H~{\sc ii} region at $s=100 \pc$, so most of the emission occurs behind most of the Faraday rotation. However, evidently the H~{\sc ii} region again does not fully block information about the ISM beyond it, even though $\sigma_\phi^2 \lambda^4 \sim 10^{2.5}$ at $\lambda=70 \, \cm$ (Figure~\ref{fig:siglambda}): the measured first moment increases from $\tphi(200 \, \pc) = +5.9 \radmsq$ to $\tphi(1 \, \kpc) = +13.5 \radmsq$ to $\tphi(3 \, \kpc) = +21.4 \radmsq$. With the moments, we lose the complexity in the Faraday spectrum. However, RM synthesis with moments still benefits from the wide $\lambda^2$ coverage, averaging over the $\sinc (\phi \lambda^2)$ behavior of depth depolarization.


Different observed Faraday depths at different wavelengths are often interpreted as evidence that the different wavelength observations are sampling different volumes of the ISM because of the moving polarization horizon. However, this picture does not describe the model presented here. In particular, as discussed at the beginning of Section~\ref{sec:results}, the sign of the rotation measure can change depending on the portion of the frequency spectrum (Figure~\ref{fig:spectrum}) being sampled in the {\tt noarm} model. There is no portion of any of the models presented here with a negative $\phi$, yet $\tphi$ can have components with a negative centroid both at HBN and LBS frequencies (Figure~\ref{fig:fdf}). Therefore, depolarization effects lead not to a Faraday spectrum that samples a different volume but measured values of $\tphi$ that do not map to $\phi$ in an obvious way.

In conclusion, I find that both discrete depolarizing structures and an extended medium with non-uniform electron density and magnetic field can create a distance at which the polarization effects begin a sunset, so the polarization horizon metaphor has clear utility. However, integrating further through the medium does continue to lead to changes in the observables: information reaches us from beyond the polarization horizon.


%

\vspace{6pt} 


\acknowledgments{I thank the organizers of the meeting ``The Power of Faraday Tomography: Towards 3D Mapping of Cosmic Magnetic Fields'' in Miyazaki, Japan from May 28--June 2, 2018; this work is based on a talk prepared for that meeting as well as discussions there, especially with C. van Eck, J. Farnes, B. M. Gaensler, N. M. McClure-Griffiths, and A. Thomson. T. L. Landecker and two anonymous referees each provided insightful comments on the manuscript which led to an improved paper. A.S.H. is partly funded by the Dunlap Institute at the University of Toronto.}


\conflictofinterests{The author declares no conflict of interest.}
\bibliographystyle{mdpi}


\bibliography{bibdesk_bibtex}


\end{document}